\documentclass[12pt,longnamesfirst,preprint]{aastex}

\slugcomment{2010 Asia-Pacific Radio Science Conference\\
in Toyama, Japan on September 22-26, 2010\\
Commission J: Radio Astronomy\\
J2 Millimeter- and Sub-millimeter-wave Telescope and Array 
}

\shorttitle{Okumura et al.}
\shortauthors{
Atacama Compact Array Correlator
}

\begin{document}


\title{
Atacama Compact Array Correlator for Atacama Large Millimeter/submillimeter Array
}


\author{
Sachiko K. Okumura\altaffilmark{1},
Yoshihiro Chikada\altaffilmark{1}, 
Takeshi Kamazaki\altaffilmark{1}, 
Takeshi Okuda\altaffilmark{2}, 
Yasutaka Kurono\altaffilmark{1}, 
and 
Satoru Iguchi\altaffilmark{1}
}  


\altaffiltext{1}{
National Astronomical Observatory of Japan, 2-21-1 Osawa,
Mitaka, Tokyo 181-8588, Japan
}
\altaffiltext{2}{
Nagoya University, Furo-cho, Chikusa-ku, Nagoya, 464-8601, Japan
}

\begin{abstract} 
We have developed a FX-architecture digital spectro-correlator, 
Atacama Compact Array Correlator for the Atacama Large 
Millimeter/submillimeter Array. 
The ACA Correlator processes four pairs of dual polarization signals, 
whose bandwidth is 2 GHz, from up to sixteen antennas, and calculates auto- 
and cross-correlation spectra including cross-polarization in all 
combinations of sixteen antennas. 
We report the detailed design of the correlator and the 
verification results of the correlator hardware.
\end{abstract}


\section{ALMA and ACA Correlator} 
ALMA (Atacama Large Millimeter/submillimeter Array) consists of fifty 
12-m antennas and ``Atacama Compact Array (ACA)''
\citep{igu09}. The ACA antennas are composed of four 12-m 
antennas for total power observation and twelve 7-m antennas for 
short-baseline interferometer. The ACA Correlator proforms the processing 
of astronomical signals from all the ACA antennas.

\section{Correlator Design and Requirements}
Astronomical correlator calculates correlation coefficients between the 
signals from antennas, and it has a function to arbitrarily change the 
integration time and frequency resolution to get the data to be handled 
easily. It also provides information on frequency spectral profiles. 
The newly developed ACA Correlator adopts ``FX'' architecture, in which the 
Fourier transform is performed before a cross multiplication. 
ALMA is a general-purpose millimeter and submillimeter telescope which can
observe all kinds of astronomical objects in the range of its angular 
resolution. There are many scientific goals, and sensitivity, frequency 
resolution, and frequency coverage are key factors to achieve the various 
scientific goals. 

In ALMA, the quantization loss is required to be smaller than 12 \%, 
which is the value of 2-bit quantization, by performing quantization with 
3-bit 4Gsps. To achieve this requirement, ACA Correlator processes the 
digital signals of 3-bit 4Gsps without reducing the number of bits. 
In addition, the correlator is designed to achieve high spectral dynamic 
range of $10000:1$. 
The ACA Correlator also supports wide variety of flexible spectral 
configurations (combinations of frequency resolution and coverage) for 
observations which span from wide continuum emission to high-resolution 
maser emissions. Major specifications of the ACA Correlator performance 
are listed in Table \ref{tbl:1}.

\section{Correlator Hardware}  
\begin{figure}[htbp]
\includegraphics[width=.95\linewidth]{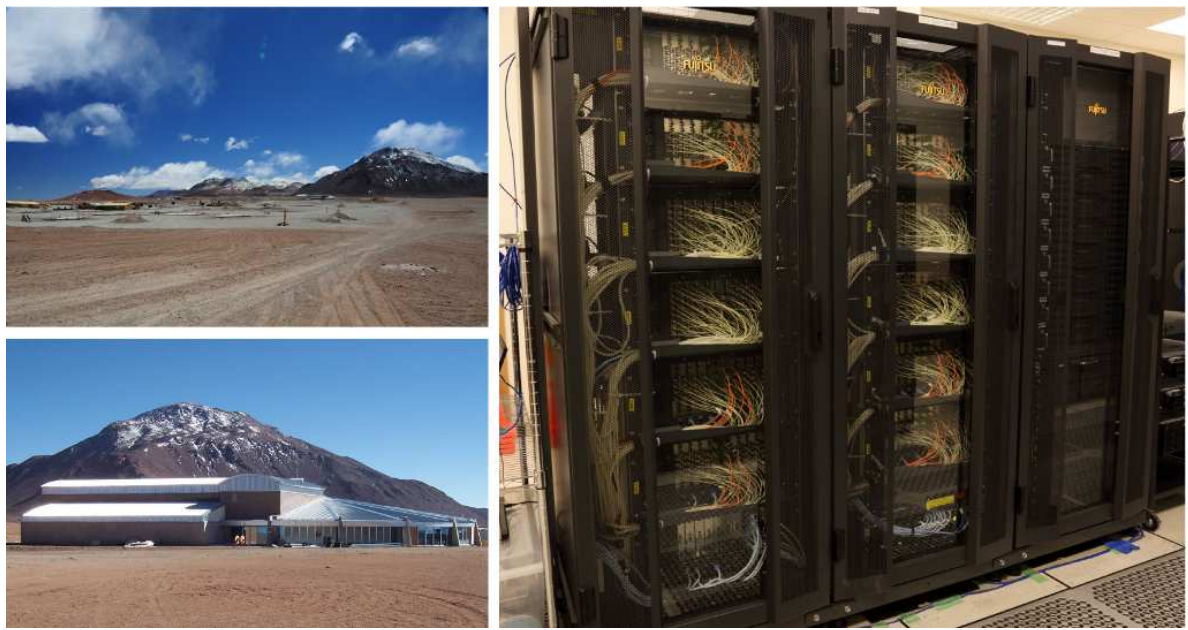}
\caption{
{\bf Upper-Left:} A photo of Array Operations Site (AOS) (September 2009).
{\bf Lower-Left:} The AOS (Array Operations Site) Technical Building. 
{\bf Right:} ACA Correlator racks.
}
\label{fig:1}
\end{figure}

The ACA Correlator was installed in the room of Array Operation Site 
Technical Building at an altitude of approximately 5000 meters 
(Figure \ref{fig:1}). 
The correlator processes 8-baseband signals from 16 antennas with 8 racks. 
Each two racks perform all the necessary processes for one pair of dual
polarization signals,
and they are composed of three kinds of box-like modules. Also, module 
structure can be easily replaced by a few persons at the high site from 
the viewpoints of operability and maintainability.

\section{Verification of the ACA Correlator}
We have performed verification tests of design and implementation 
for the scientific function, stability, and performance of the ACA 
Correlator.

\subsection{Function test using known signal data}
\begin{figure}[htbp]
\includegraphics[width=\linewidth]{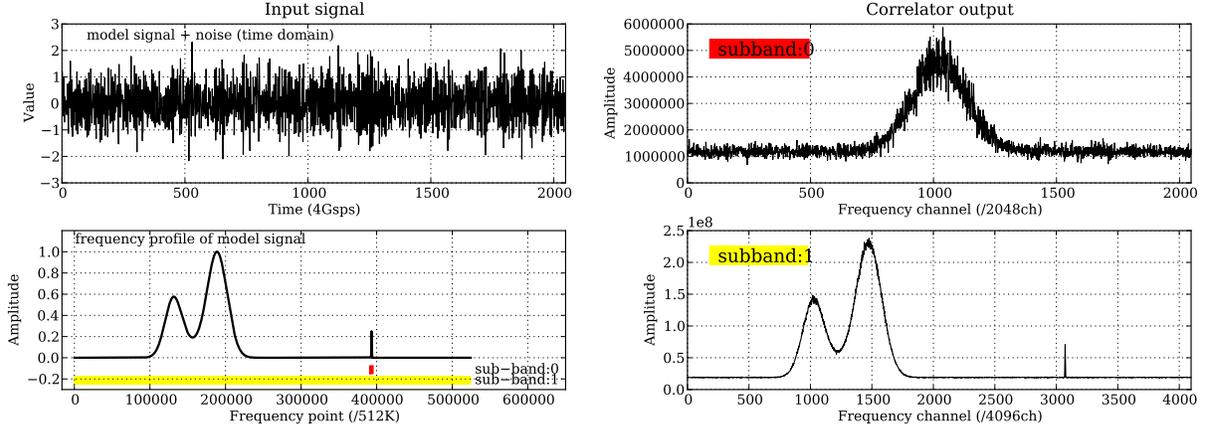}
\caption{
{\bf Left:} An example of input data. For this input, the ACA Correlator 
was configured to have two sub-bands within the 2GHz baseband; one has a 
bandwidth of 15.625MHz with a frequency resolution of 7.629kHz (Red), and 
the other has a bandwidth of 2GHz with a frequency resolution of 488.28kHz 
(Yellow).
{\bf Right:} Correlator output (autocorrelation) for the sub-band 0 and 1, 
which show the autocorrelation spectra of a part of 2GHz bandwidth with 
high-resolution and overall profile within the 2GHz baseband.
}
\label{fig:2}
\end{figure}

The output of calculated results from the ACA Correlator were checked to be
bit-accurately coincident with our developed simulator using known input
data, for various sets of correlator configurations.
Figure \ref{fig:2} demonstrates an example of verification tests of ACA 
Correlator using known data.

\subsection{Long-term integration tests of analog noise}
\begin{figure}[htbp]
\begin{minipage}{0.58\linewidth}
\includegraphics[width=\linewidth]{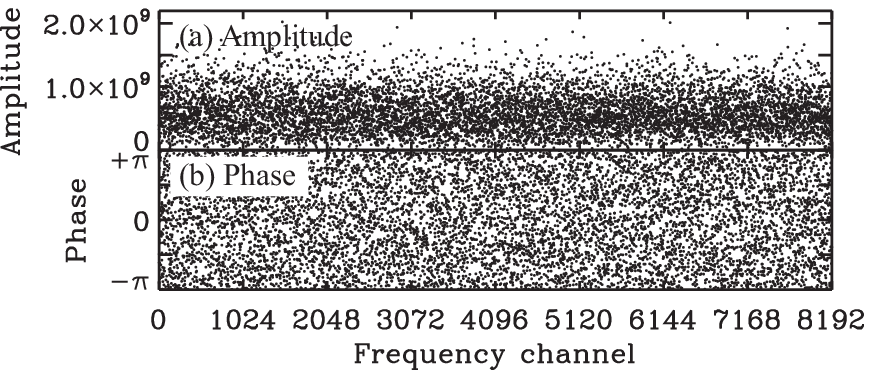}
\end{minipage}
\hspace{.5cm}
\begin{minipage}{0.4\linewidth}
\includegraphics[width=\linewidth]{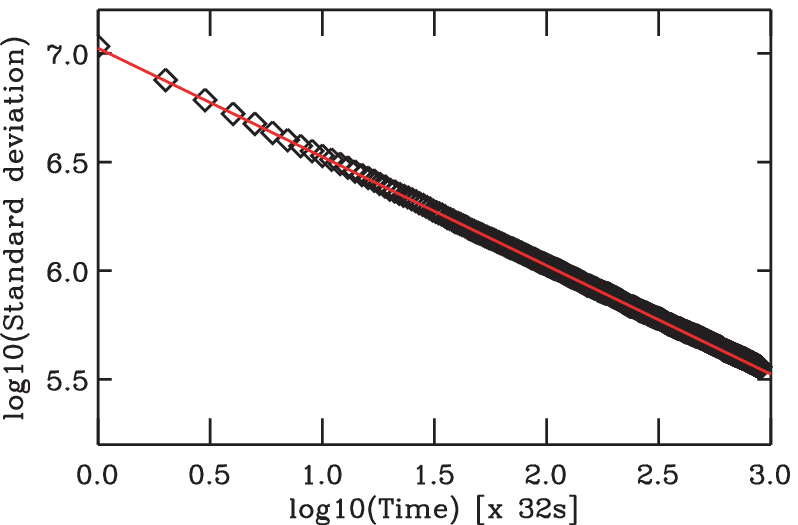}
\end{minipage}
\vspace{.5cm}
\caption{
{\bf Left:} 
Amplitude and phase of cross-correlation spectra integrated over eight
hours. Its spectral bandwidth and resolution are 31.25 MHz and 3.815
kHz, respectively.
{\bf Right:} 
Amplitude root-mean-square of cross-correlation spectra as a function
of the total integration time. It is overlaid by a red line with
inclination of $1/\sqrt{\rm integration~ time}$. 
}
\label{fig:3}
\end{figure}

We conducted long-term integration test of analog noise in order
to verify the performance of ACA Correlator. In this test, we
verified the following two items by inputting 1-bit digitized thermal
noise to the correlator and integrating the cross-correlation products
for 8 hours:
\begin{itemize}
\item There is no harmful sporadic event caused by digital processing
\item Noise level decrease in proportion to $1/\sqrt{\rm integration~ time}$
\end{itemize}
Figure \ref{fig:3} shows this test results, which confirm the two
verification items;
The left plot clearly shows there is no artificial
pattern caused by digital processing, and for the right plot
it can be clearly seen that the noise level decreases in proportion to 
$1/\sqrt{\rm integration~ time}$.

\subsection{Signal receiving and data processing from ALMA antennas}
\begin{figure}[htbp]
\begin{minipage}{0.51\linewidth}
\includegraphics[width=\linewidth]{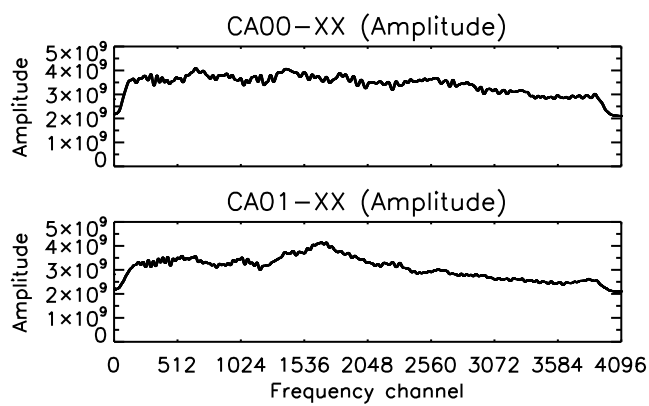}
\end{minipage}
\hspace{.5cm}
\begin{minipage}{0.45\linewidth}
\includegraphics[width=\linewidth]{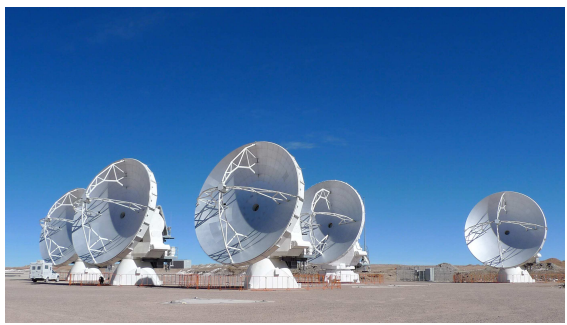}
\end{minipage}
\vspace{.75cm}
\caption{
{\bf Left:} ACA Correlator output for antenna input CA00 and CA01: 
amplitude of auto-correlation spectra for the signal from two ALMA
antennas (Quadrant-1, bandwidth:2GHz, resolution:488kHz, 4096 frequency channels/polarization, polarization:XX, integration:4s).
{\bf Right:} ALMA 12-m antennas at AOS.
}
\label{fig:4}
\end{figure}

We also conducted the test of receiving signals from ALMA antennas
and data processing at Array Operation Site (AOS), and verified that the 
ACA Correlator successfully received the signals and processed the 
digitized data without any problems (Figure \ref{fig:4}).

\begin{table}[htbp]
\caption{Major specifications of the ACA Correlator}
\label{tbl:1}
\begin{tabular}{p{14em} c p{21em}}
\hline
 Function & \hspace{3pt} & Specification\\
 \hline
 Number of antennas &  & 16\\
 Number of inputs per antenna &  & 4 baseband $\times$ 2 polarizations $=$ 8 baseband\\
 Processing bandwidth per input &  & 2 GHz\\
 Number of bits per sample &  & 3 bits\\
 Maximum delay compensation	&  & $>$ 18.5 km\\
 Maximum number of processed correlations per input &  & 120 cross-correlations $+$ 16 auto-correlations\\
 Highest frequency resolution &  & 3.8 kHz\\
 Integration in the time domain &  & 1 or 16 msec (for auto-correlations), 16 msec (for cross-correlations)\\
 Maximum correlator output rate &  & 8192 frequency channels per baseband\\
 \hline
\end{tabular}
\end{table}

\end{document}